\documentclass{aa}
\usepackage{psfig}
\begin{document}
   \thesaurus{1         
              (11.09.1 LRG J0239--0134 ; 
               11.05.2;  
               11.09.2;  
               11.19.1;  
               12.03.3;  
               12.07.1)} 

\title{A ring galaxy at $z=1$ lensed by the cluster Abell 370
\thanks{Based on observations with the NASA/ESA {\it Hubble Space
Telescope\/} obtained from the data archive at the Space Telescope
European Coordinating Facility, with {\it ISO}; an ESA project with
instruments funded by ESA Member States with the participation of ISAS
and NASA, and with the Canada-France-Hawaii Telescope at Mauna Kea,
Hawaii, USA.}  }

   \author{
G. Soucail\inst{1} \and
J.P. Kneib\inst{1} \and
J. B\'ezecourt\inst{1, 2} \and
L. Metcalfe\inst{3} \and
B. Altieri\inst{3} \and
J.F. Le Borgne\inst{1}
   }

   \offprints{G. Soucail, soucail@obs-mip.fr}

   \institute{Observatoire Midi-Pyr\'en\'ees, Laboratoire d'Astrophysique, 
    UMR 5572, 14 Avenue E. Belin, F-31400 Toulouse, France \and
    Kapteyn Institute, Postbus 800, 9700 AV Groningen, The Netherlands
    \and {\it ISO} Data Centre, Astrophysics Division, Space Science
    Department of ESA, Villafranca del Castillo, PO Box 50727, E-28080
    Madrid, Spain
   }

\date{Received 12 January 1999 / Accepted 26 January 1999}

\maketitle

\markboth{G. Soucail et al.: A ring galaxy lensed by A370}{}

\begin{abstract} We present a study of a very peculiar object found in
the field of the cluster-lens Abell 370. This object displays, in {\it
HST} imaging, a spectacular morphology comparable to nearby
ring-galaxies.  From spectroscopic observations at the CFHT, we
measured a redshift of $z=1.062$ based on the identification of [O{\sc
ii}] 3727\AA\ and [Ne{\sc v}] 3426\AA\ emission lines.  These emission
lines are typical of starburst galaxies hosting a central active
nucleus and are in good agreement with the assumption that this object
is a ring galaxy. This object is also detected with {\it ISO} in the
LW2 and LW3 filters, and the mid Infra-Red (MIR) flux ratio favors a
Seyfert 1 type.  The shape of the ring is gravitationally distorted by
the cluster-lens, and in particular by a nearby cluster elliptical
galaxy. Using our cluster mass model, we can compute its intrinsic
shape. Requiring that the outer ring follows an ellipse we constrain
the M/L ratio of the nearby galaxy and derive a magnification
factor of 2.5 $\pm$ 0.2. The absolute luminosities of the source are
then $L_B = 1.3 \ 10^{12} L_{B \odot}$ and $\nu$ L$_\nu \simeq
4. 10^{10}$ L$_\odot$ in the mid-IR.

\end{abstract}

\section{Introduction} A useful property of gravitational lensing is
the magnification of background objects: the gain in spatial
resolution allows the morphological properties of distant and resolved
objects to be probed in greater detail and the gain in apparent flux
allows fainter sources than would otherwise either be detected or
studied to be probed statistically. Massive clusters of galaxies 
can then be used as natural ``gravitational telescopes'' to address
several astrophysical problems related to the properties and nature of
high redshift galaxies.  Morphological properties of distant lensed
sources were first addressed by Smail et al.\ (1996) who estimated the
intrinsic linear sizes of the galaxies and showed them to be compatible
with a significant size evolution with redshift.  More recently,
several detailed morphological analysis of high-$z$ arcs were proposed
using lens modeling. Most of them appear knotty with a more complex
morphology than their local counterparts (\cite{colley96};
\cite{franx97}; \cite{pello99}) although there may be some biases due
to observations in the UV restframe.

In this letter, we study a peculiar object detected in the field of
the galaxy cluster A370. On the {\it HST}/{\it WFPC2\/} images, it
displays an unusual morphology similar to nearby ring-galaxies,
although gravitationally distorted by the cluster.  Sect. 2 summarises
the various observations relating to this source: {\it HST} imaging,
spectroscopic and photometric data including mid-IR photometry
obtained with ESA's {\it ISO} spacecraft (\cite{kessler96}). The
source reconstruction of the ring and a morphological analysis is
presented in Sect. 3 with a robust estimate of the lens amplification.
Sect. 4 discusses the spectral energy distribution (SED) of the
object. Discussion and conclusions are given in the last section.

Throughout the paper, we consider a Hubble constant of H$_0 = 50 \
h_{50}$ km s$^{-1}$ Mpc$^{-1}$, with $\Lambda = 0$ and $\Omega=1$.

\section{Observations}
\subsection{{\it HST} imaging}
\begin{figure*}
\centerline{
\psfig{figure=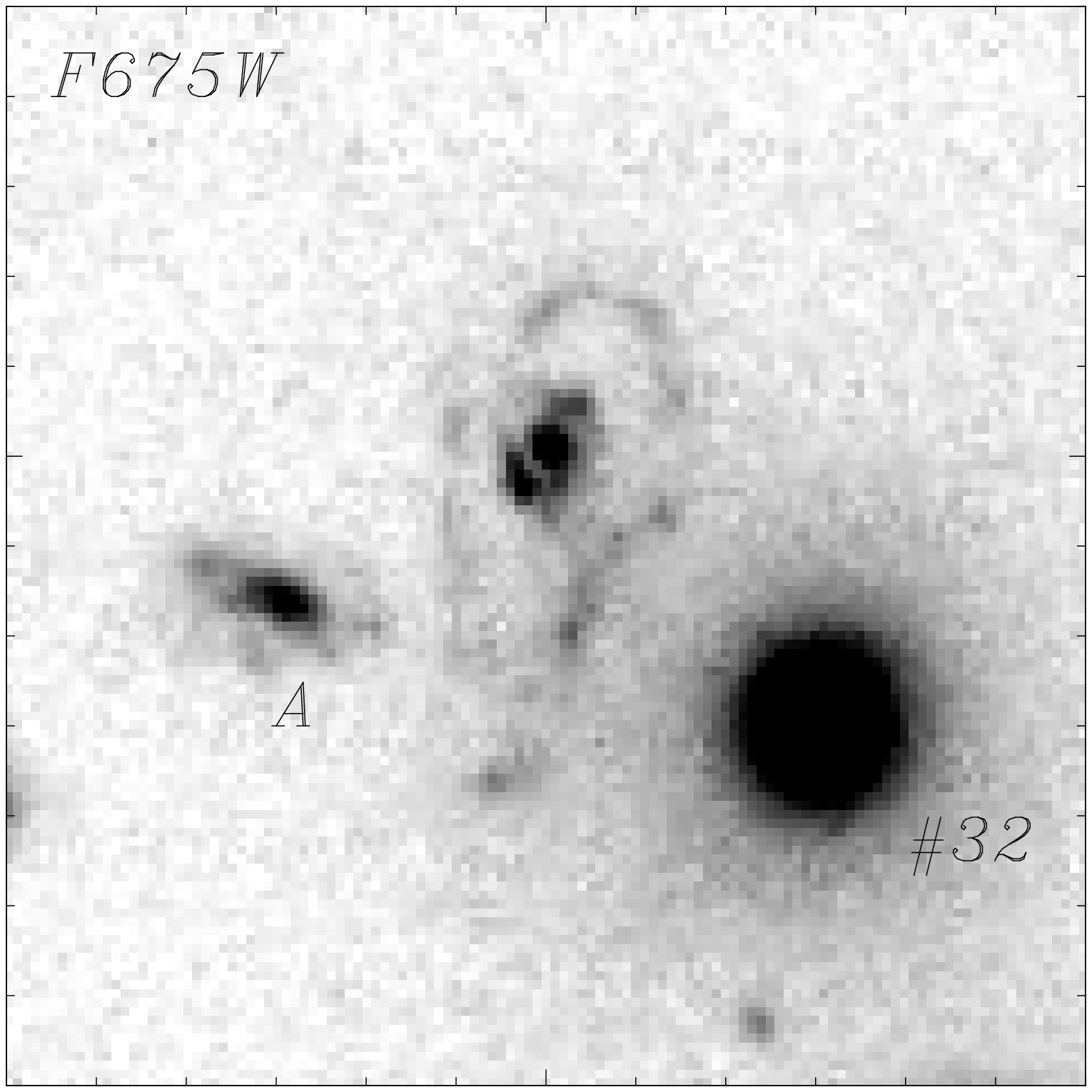,width=0.33\textwidth}
\psfig{figure=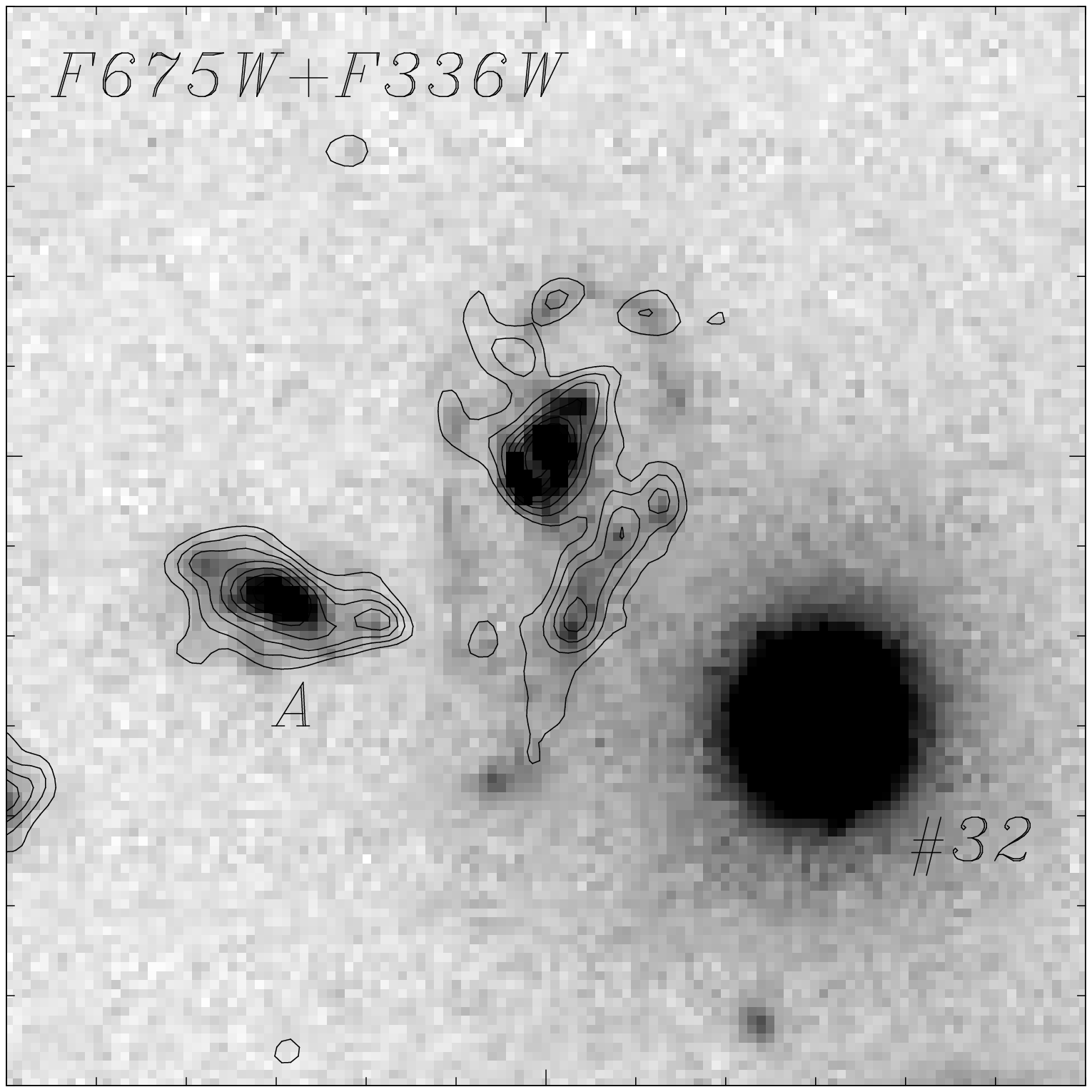,width=0.33\textwidth}
\psfig{figure=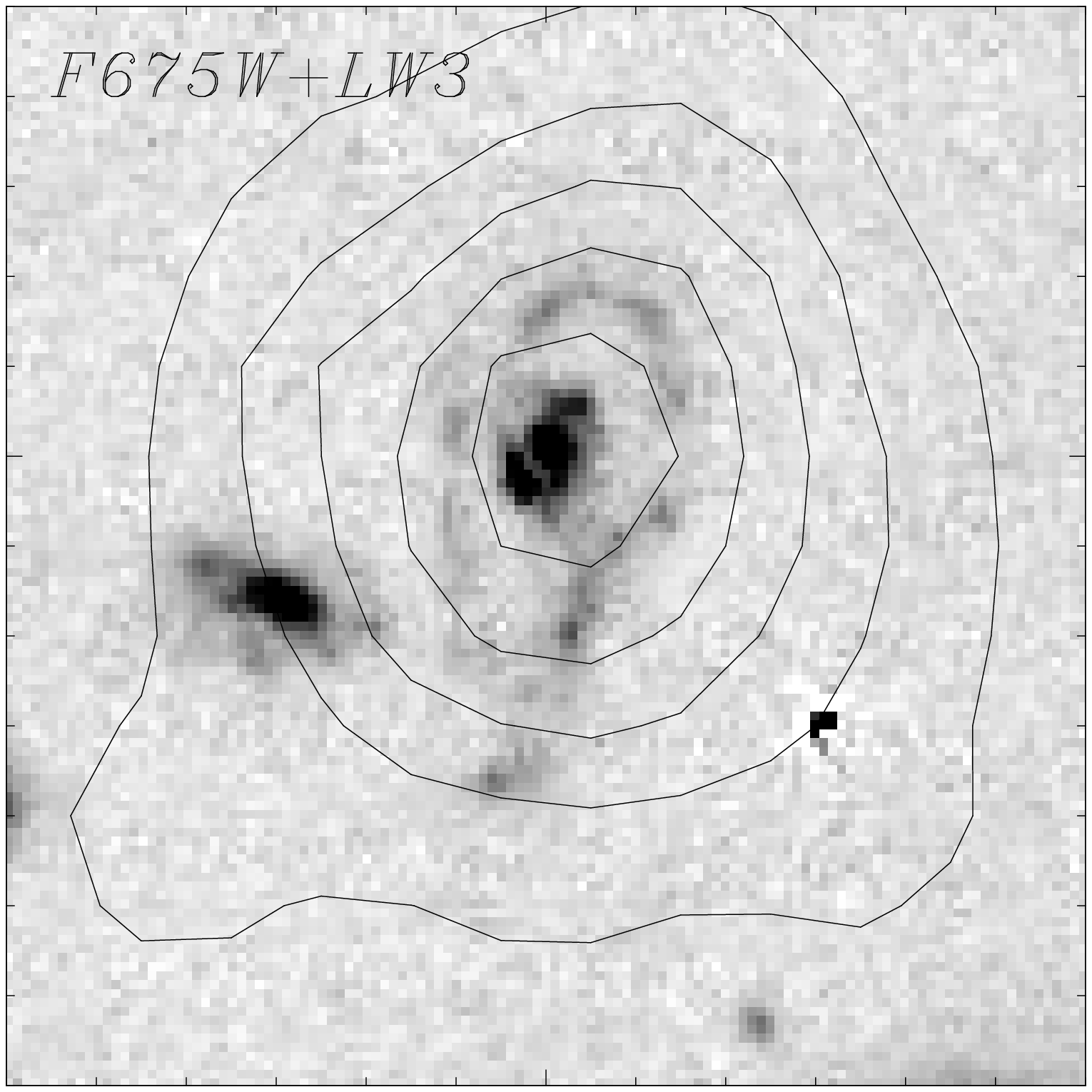,width=0.33\textwidth}
}
\caption{Multi-colour images of the ring galaxy and its
surroundings: {\it Left:} {\it HST}{\it WFPC2\/} F675W image
(with a 12\arcsec\ size) -- {\it Center:} isocontours of the {\it HST} 
F336W (U-band) image overplotted on the F675W image -- {\it Right:}
{\it ISO} LW3 contours overplotted on the F675W image. Galaxy
\# 32 has been subtracted for clarity. Note the accurate matching of
the {\it ISO} source with the nucleus. North is top, East is left.}
\label{fig-ring}
\end{figure*}

In the deep F675W {\it WFPC2\/} image described in B\'ezecourt
et al.\ (1999a, hereafter Paper I), a spectacular distorted ring is
detected close to a bright cluster elliptical (\#32 in the numbering
scheme of Mellier et al.\ (1988) with $z=0.370$).  The object displays a
clearly resolved central bulge surrounded by a $\sim$4.8\arcsec-diameter
distorted ring and a secondary 1.5\arcsec-diameter inner ring (Figure
\ref{fig-ring}).  This object is very similar in aspect to the Cartwheel
galaxy (Struck et al.\ 1996), with similar ratio between the two ring
radii.  We used the APM catalogue to measure the astrometry of the
field.  The absolute coordinates of the lensed ring galaxy are:
$\alpha_{J2000} = $ 2h 39m 56.51s, $\delta_{J2000} = -1^\circ$
34\arcmin\ 25.66\arcsec\ (with a 0.2\arcsec\ rms accuracy). Therefore
we reference hereafter this object as: LRG J0239--0134.

A photometric analysis was performed on the F675W {\it HST} image.  First,
the nearby galaxy \# 32 was subtracted after a radial fit of the isophotes
with the ``ELLIPSE'' package in the IRAF/STSDAS environment. The total
integrated flux for LRG J0239--0134 gives a magnitude R$_{675W} = 20.7
\pm 0.1$ that can be separated into 
the emission of the central part R$_{675W} = 21.5 \pm 0.1$ and the
outer ring contribution R$_{675W} = 21.4 \pm 0.2$ (less accurate due
to larger uncertainties in the local sky background).  In the {\it
HST}/{\it WFPC2\/} U-Band image (see B\'ezecourt et al.\ 1999b) the
ring-like object is also detected and appears less centrally
concentrated than in F675W. However the lower signal-to-noise prevents
a detailed morphological analysis of the extended emission (Figure
\ref{fig-ring}) although there may be some inhomogeneities in the ring
due to hot spots similar to that seen in the Cartwheel.

\subsection{Spectroscopic and photometric observations}
\begin{figure}
\centerline{\psfig{figure=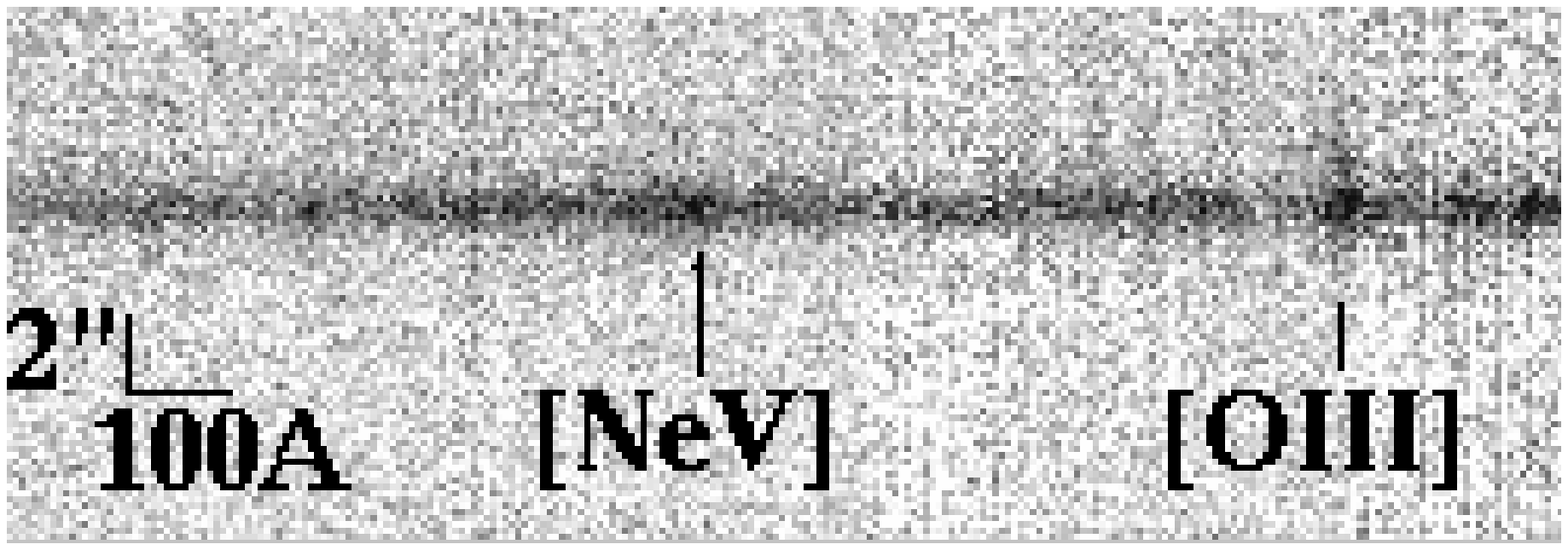,width=0.3\textwidth}}
\psfig{figure=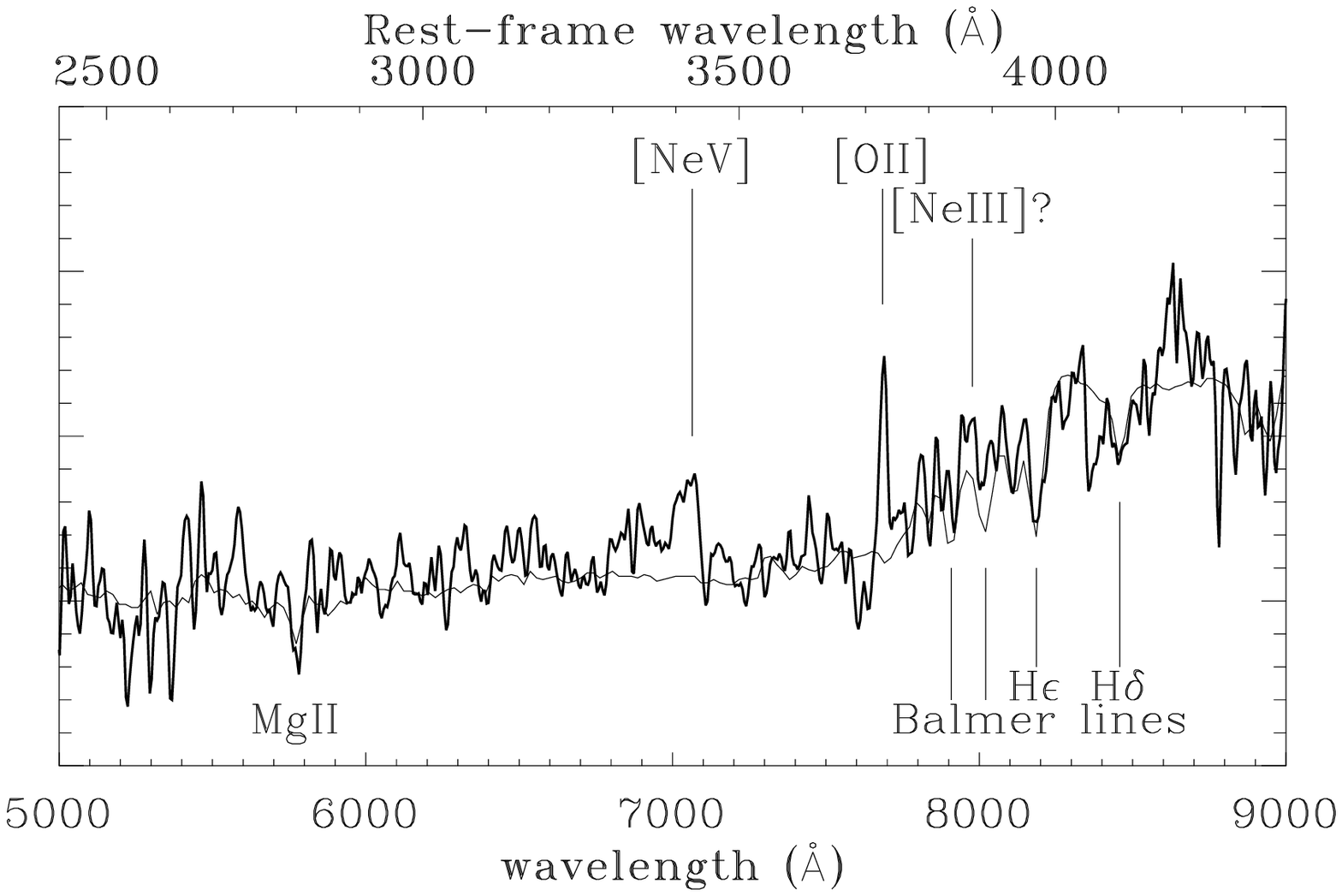,width=0.5\textwidth}
\caption{The spectrum of the ring galaxy superimposed on a synthetic
spectrum of a late-type spiral redshifted to $z=1.062$. The data were smoothed 
to fit the resolution of the spectrograph and the most
prominent lines are identified. Above is the original 2D spectrum
after sky subtraction, showing in particular the spatial extent of the
emission line at 7685\AA.}
\label{spec-ring}
\end{figure}

Spectroscopic data were obtained during a CFHT run with the OSIS-V
instrument (\cite{lefevre94}) in August 1997. A 1{\arcsec}-wide
long-slit was positioned through the central bulge and the ring 
(slit orientation was East-West limiting the contamination by
the envelope of galaxy \#32). We used the
2K$\times$2K Loral thinned CCD and the grism R150 which gives a
dispersion of 6\AA/pixel, with a resolution of 18\AA\ across the
wavelength range 5000 to 9000\AA. Two 1.8 ksec exposures were obtained
just before morning twilight. 
The data were reduced with standard procedures for flat-fielding,
wavelength and flux calibration. Sky subtraction was performed on the
2D image (Figure \ref{spec-ring}). Several features are detected: a
strong emission line at $\lambda = 7685 \pm 3$\AA, a weaker one at
$\lambda = 7064 \pm 9$\AA\ and an absorption line at $\lambda = 5773
\pm 3$\AA, all visible on the 2D spectrum.  We unambiguously identify
those lines as: [O{\sc ii}] 3727\AA\,,[Ne{\sc v}] 3426\AA\ and Mg{\sc
ii} 2800\AA\ giving a redshift of $z=1.062 \pm 0.001$ for the source.
Moreover, a tentative identification at this redshift suggests weak
emission from the [Ne{\sc iii}] 3869\AA\ line though the spectrum is
contaminated by sky residuals and H$\delta$ and H$\varepsilon$
absorption lines. Redder lines fall outside the visible range and
unluckily, H$\alpha$ lies at 1.353$\mu$m, a wavelength domain hardly
detectable from the ground (outside the J-band). Furthermore, the fit
of the continuum with a typical late-type spiral galaxy is
satisfactory. Finally, the [O{\sc ii}] line is more spatially extended
than the underlying stellar continuum, and a contribution from the
ring itself is likely, which is not the case for the [Ne{\sc v}]
line. Unfortunately, no information on any velocity field can be
provided with this low S/N ratio spectrum.

Multi-band photometry of this galaxy is available from a large set
of data existing on the cluster Abell 370. In addition to the {\it
HST} U and R images, we used B, R and I magnitudes from deep CFHT
images (\cite{kneib94}). Near-IR data were obtained with the
Redeye camera at CFHT in August 1994. The final near-IR images
correspond to deep exposures with integration times of 3.4 ksec in
J and 7.2 ksec in K$'$ in good observing conditions. The details
of this photometric setup will be published elsewhere. The
photometry of the ring galaxy is summarised in Table
\ref{mag-multil}. From this multi-band photometry, a photometric
redshift of $0.95 \pm 0.1$ was found (R. Pell\'o, private
communication) confirming the spectroscopic identification through a
fit to the continuum shape.  In addition, the fainter object close to
the ring (labelled A in Figure \ref{fig-ring}) has a photometric
redshift $z_{phot} = 0.4 \pm 0.1$ and is therefore not related to the
ring galaxy.

\begin{table}
\caption[]{Multi-wavelength fluxes of the ring galaxy (see text for more
details). These fluxes are not corrected for the gravitational
magnification of the source.}
\label{mag-multil}
\begin{flushleft}
\begin{tabular}{lllcc}
\hline\noalign{\smallskip}
Filter & $\lambda$ ($\mu$m) & Magnitude & Flux ($\mu$Jy) & Resolution \\
\noalign{\smallskip}
\hline\noalign{\smallskip}
U$_{336W}$ & 0.336 & 21.00 $\pm 0.2$ & $4.45 \pm 0.9$ & 0.1 \arcsec\ \\
B$_j$ & 0.450 & 21.81 $\pm 0.2$ & $6.36 \pm 1.2$ & 1.1 \arcsec\ \\
R & 0.646 & 20.38 $\pm 0.4$ & $21.3 \pm 9.4$ & 1.1 \arcsec\ \\
R$_{675W}$ & 0.673 & 20.66 $\pm 0.1$ & $15.8 \pm 1.5$ & 0.1 \arcsec\ \\
I & 0.813 & 19.31 $\pm 0.2$ & $45 \pm 9$ & 0.7 \arcsec\ \\
J & 1.237 & 17.82 $\pm 0.2$ & $118 \pm 24$ & 1.2 \arcsec\ \\
K$'$ & 2.103 & 16.34 $\pm 0.2$ & $212 \pm 42$ & 1.2 \arcsec\ \\
LW2 & 6.7 & & $750 \pm 225$ & 6 \arcsec\ \\
LW3 & 14.3 & & $1800 \pm 540$ & 6 \arcsec\ \\
\noalign{\smallskip}
\hline
\end{tabular}
\end{flushleft}
\end{table}

\subsection{{\it ISO} data} The ring galaxy was detected with the
{\it ISO} camera ({\it ISOCAM}, C\'esarsky et al., 1996), as part
of a programme of imaging through gravitational lensing clusters
(Metcalfe et al.\ 1999). A370 was deeply imaged on a wide $7'
\times 7'$ field, in micro-scanning mode, with the 3\arcsec\
pixel-field-of-view, and using two broad-band, high-sensitivity
filters: LW2 (5--8.5$\mu$m) and LW3 (12--18$\mu$m) with a total
exposure time of 16.1 ksec in the deepest part of the image.  The
data were reduced following two substantially independent methods: a
Multi-resolution Median Transform method (PRETI, Starck et al.\
1998), and the Vilspa method described in Altieri et al.\
(1998). The centering of {\it ISO} maps on optical data was obtained
by correlating the positions of 7 sources, giving a final
astrometric accuracy of 1\arcsec . The correspondance between the
ring galaxy and the {\it ISO} source is quite secure (Figure
\ref{fig-ring}). The photometric accuracy achievable for such faint
mid-IR sources, allowing for all the uncertainties in the data
reduction, is around 30\%. In both {\it ISO} bandpasses, the ring
galaxy is the brightest extragalactic source in the field covered by
{\it ISOCAM}, apart from the giant arc.

\section{Source reconstruction of the lensed ring galaxy}
\begin{figure}
\centerline{\psfig{figure=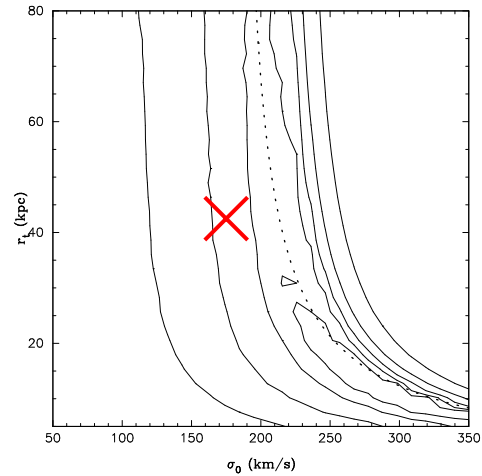,width=0.4\textwidth}}
\caption{Contour plot of the estimator E for 
the galaxy \#32 when fitting the outer ring to be projected as an
ellipse in the source plane.  The dotted line correponds to models
with constant M($<a_{32}$)/L. 
}
\label{ring_ml}
\end{figure}

This ring galaxy is magnified and distorted by the gravitational
shear induced by the cluster and the nearby elliptical cluster galaxy
\#32.  This is particularly visible in the distorted shape of the
outer ring.  In order to reconstruct its true intrinsic size, shape and
determine its total amplification, we traced the rays back through the
lens into the source plane, using the model proposed in Paper I. This
model was optimised from the identification of several multiple images
detected on the {\it HST} images and takes into account the massive
halos of the brightest galaxies using standard scaling laws ({\it e.g.}
\cite{kneib96}) related to their luminosities with scaling dependance
similar to the properties of the Fundamental Plane: 
$$
\sigma_0=\sigma_{0\ast} \left({L\over L_{\ast}}\right)^{{1\over 4}};
\ \ \,
r_t=r_{t\ast} \left({L\over L_{\ast}}\right)^{{0.8}};
\ \ \,
r_0=r_{0\ast} \left({L\over L_{\ast}}\right)^{{1\over 2}}.
$$

In particular for the elliptical galaxy \#32 (R$_{675W} = 18.47$) this
means that L$_R =2.2 \ 10^{11} \ h_{50}^{-2}$ L$_\odot$ which
translates to $\sigma_0 =$ 173 km/s, $r_t =$ 42.5 $h_{50}^{-1}$ kpc (or
6.8\arcsec\ at the cluster redshift) and a total M/L ratio of 6.4
$h_{50}$ (M/L)$_\odot$ for this galaxy and its halo. Within the
aperture $a_{32}=26.8 \ h_{50}^{-1} $ kpc defined by the distance from
\#32 to the ring galaxy nucleus we have M($<a_{32}$)/L = 5.7 $h_{50}$
(M/L)$_\odot$.

\begin{figure}
\centerline{
\psfig{figure=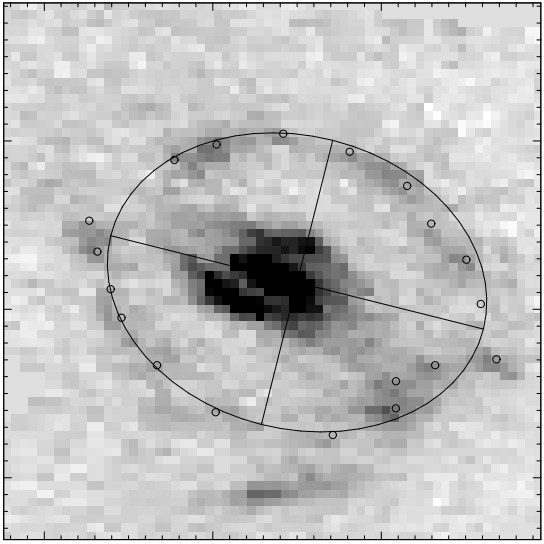,width=0.4\textwidth}
}
\caption{Image-reconstruction of the ring galaxy in the source
plane using the best fit model parameters. The bottom part of the ring
was not included in the fitting. This reconstruction suggests that
this object does not belong to the outer ring.}
\label{sring} \end{figure}

We then examined the influence of the parameters ($\sigma_0, r_t$) in
the source reconstruction.  In order to quantify this we identified in
the image plane 20 points ($x_i,y_i$) describing the outer ring. We
then tuned the two parameters ($\sigma_0, r_t$) to check how close to
an ellipse the corresponding source points ($xs_i,ys_i$) are. For this
purpose we defined the following estimator:
\[
E(\sigma_0, r_t)={1\over N}\Sigma_{i=1}^N | f_{\sigma_0, r_t}(i) |
\]
with 
\begin{eqnarray}
f_{\sigma_0, r_t}(i)&= &
{{[(xs_i-x_c)\cos \theta - (ys_i-y_c)\cos \theta ] ^2}
        \over a^2} \nonumber \\
& + & {{[(xs_i-x_c)\sin \theta + (ys_i-y_c)\sin \theta ] ^2}
        \over b^2} - 1 \nonumber
\end{eqnarray} where ($x_c, y_c, a, b, \theta$) are the parameters of
the ellipse that minimise, for each set of ($\sigma_0, r_t$), the
estimator E.  For a set of points belonging to an ellipse, E is
zero. Furthermore, by definition, E is scale-invariant and does not
depend on the intrinsic size of the ring.  Figure 4 shows iso-contours
of the estimator E($\sigma_0, r_t$). The original point of the model
of paper I is indicated as a cross. Not surprinsingly, the shape of
the ring does depend strongly on the M/L ratio within the aperture
$a_{32}$.  Fixing $r_t$=42.5 $h_{50}^{-1}$kpc, the best value for the
velocity dispersion is $\sigma_0 =$ 220 $\pm 20$ km/s, which
corresponds to a correction factor of 1.4 for the aperture mass
leading to M($<a_{32}$)/L = 8.0 $\pm 1.5$ (M/L)$_\odot$.

The magnification factor of the source has a mean value of 2.5$\pm
0.2$ but ranges from 3.6 to 2.1 depending on the distance to galaxy
\#32 (Figure
\ref{sring}). The intrinsic radius for the outer ring found is $7.7
\pm 0.4 \, h_{50}^{-1}$ kpc, a value comparable with the
characteristics of nearby similar objects. The axis ratio of the
ellipse is $\cos i = b/a =0.76$ that translates in a inclination angle
of $i=$40 degree assuming an intrinsic circular ring.  The B-band
absolute luminosity of the source, corrected from the magnification and
measured directly from the I magnitude, is $L_B = 1.3 \ 10^{12} L_{B
\odot}$, about 10 times brighter than the Cartwheel galaxy (Appleton \&
Marston 1997).

\section{Spectral Energy Distribution }
\begin{figure}
\psfig{figure=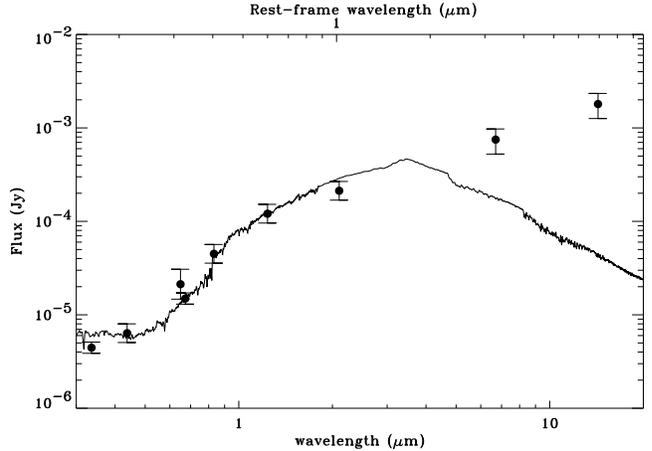,width=0.5\textwidth}
\caption{The spectral energy distribution of the ring galaxy. The data
correspond to the observed photometry from 0.3 $\mu$m to 15 $\mu$m. The
curve shows the synthetic SED of an Sc galaxy from Fioc and
Rocca-Volmerange (1996). The mid-IR excess is clearly non-stellar.}
\label{sed}
\end{figure}

Joining together all the multi-wavelength photometric measurements, we
can construct the SED of the ring galaxy over a large wavelength range
(Fig.~\ref{sed}).  This range covers both the rest-frame stellar
emission and the IR flux emitted by the warm component of the
interstellar medium. The significant excess of light emitted in the
mid-IR with respect to the stellar contribution can originate from
three possible sources: either a central active nucleus heats the
dust torus around it, or the warm dust is heated by a violent
starburst induced by the merger that produced the ring, or we are
witnessing a combination of the two phenomena.  A strong argument in
favour of an active nucleus is the detection in the optical spectrum
of the [Ne{\sc v}] emission line typical of Seyfert galaxies.
Moreover, the bulk of the optical emission is clearly stellar, with
Balmer absorption lines and a typical stellar continuum. This suggests
that a starburst is occurring and may dominate the optical light.  For
this target in the mid-IR it is difficult to differentiate between the
contribution from the nuclear emission and that from the starburst
(see the discussion of the Cartwheel galaxy in
\cite{charmandaris98}). The flux ratio LW3/LW2 is about 2.4, 
or equivalently the spectral index is about $-1.15$ for LRG
J0239--1034. Taking into account the fact that we observe the
rest-frame fluxes at 3.3$\mu$m and 7$\mu$m, this can be compared
easily with the mid-IR spectra shown by Schultz et al.\ (1998) for
different types of AGN. Our results clearly favour a Seyfert 1 type,
in accordance with the weak detection of the [Ne{\sc v}] emission line
in the optical spectrum.

\section{Discussion} 
In this letter we analyse in details the nature of a pecular galaxy
based on multiwavelength observations. 
Firstly, the outer ring of the galaxy constitutes good
morphological evidence for a starburst induced by a recent
gravitational interaction, although the progenitor is not yet
identified.  The presence of an extended [O{\sc ii}] 3727\AA\ emission
line in the 2D spectrum demonstrates the occurrence of star formation
in the ring. In addition, the galaxy hosts an active nucleus in its
centre, probably triggered by the interaction, as shown both by the
detection of the [Ne{\sc v}] line in the optical and by the properties
of the mid-IR fluxes. This ring galaxy also corresponds to the source
L3 detected in the sub-millimeter domain with SCUBA by Smail et al.\
(1998). This shows that emission arises from cold dust in the source,
but without adequate spatial information no conclusion can be drawn
about the preferred location of this cold dust (in the ring or
nucleus). We can also compute its intrinsic mid-IR luminosity, after
correction for both the gravitational magnification and a k-correction
estimated from a power-law fit to the spectrum. This gives $F (3.3 \mu
m) = 1.7 \, 10^{23}$ W/Hz and $F (7 \mu m) = 3.7 \, 10^{23}$ W/Hz, or
equivalently in solar luminosity: $\nu$ L$_\nu \simeq
4. 10^{10}$ L$_\odot$. In comparison, the total mid-IR luminosity 
of the Cartwheel at similar wavelengths is only  $\nu$ L$_\nu \simeq
2. 10^{8}$ L$_\odot$. The difference in luminosity comes most probably 
by the nucleus emission, relatively weak in the Cartwheel. LRG
J0239--0134 belongs more specifically to the sample of high redshift
galaxies for which the mid-IR and sub-mm fluxes emissions are powered
by a strong and obscured active nucleus (\cite{ivison98}), possibly
triggered by galaxy interactions.

In conclusion, we have shown in this letter that gravitational
magnification can be used to give a detailed analysis of the nature of
a few peculiar distant galaxies.  Clearly cluster-lenses are and will
be useful in the future to observe and study the nature of the most
distant galaxies in a wide domain of wavelengths and emission
processes.

\acknowledgements We wish to thank Roser Pell\'o for her help in the
estimations of photometric redshifts of the objects and Dani\`ele
Alloin for fruitful discussions about AGN optical spectra. Many thanks
to Ian Smail, Rob Ivison, Andrew Blain and Jim Higdon for useful
discussions and comments.  This research has been partly conducted
under the auspices of a European TMR network programme made possible
via generous financial support from the European Commission \\ 
{\tt (http://www.ast.cam.ac.uk/IoA/lensnet/)}.

\end{document}